# PHASE TRANSITION IN SIMPLEST PLASMA MODELS

I.L. Iosilevski, A.Yu. Chigvintsev

*Moscow Institute of Physics and Technology Dolgoprudny, Moscow Region, 141700, Russia*

Abstract

Previous study of properties of the first-order phase transition in a set of plasma models with common feature - absence of individual correlations between charges of opposite sign, was continued. Predicted discontinuities in equilibrium non-uniform charge profiles have been calculated for the case of electron distribution in atomic cell and some other similar situations. These anomalies appear when some simple approximations are used and they may be naturally interpreted in terms of discussed phase transition.

## 1.Introduction

Problem of Plasma Phase Transition (PPT) is of great interest in plasma theory during very long time [1-3]. In our previous study [4-6] and in present work we deal with a set of simplified plasma models with common feature - absence of individual correlations between charges of opposite sign. Well-known example of the model is One Component Plasma (OCP) in a *rigid*, uniform, compensating background [7,8]. Subjects of our interest:
- The OCP in a *compressible* but still uniform background
- Superposition of two *compressible* OCP of charged particles of opposite sign.

Main simplification that we suppose to be valid for these variants of the OCP is that the total equation of state (EQS) is a sum of EQS of both subsystems, and if we know these EQS, we can establish the existence of phase transitions in the model and calculate its parameters. In present work we studied the next models:

**Is**) Classical Point Charges + background of ideal fermi-gas of electrons ("Simple-OCP")
**IIs**) Quantum Interacting Electron Gas + positive electrostatic background
**IIIs**) Classical Charged Hard Spheres + electrostatic compensating background
**Id**) Non-symmetric "Double-OCP" model - Superposition of two OCP of positive and negative charges with different masses (for example, electron - proton system).

In our calculations we used the well-known analytical fits for equation of state of subsystems: fluid [7] and crystal [9] phases of OCP of ions; background of ideal fermi-gas of electrons [10], quantum interacting electron gas [8], fluid phase of classical charged hard spheres [11] etc.

## 2. Phase coexistence in the Electron gas and in the Double-OCP model

We have calculated parameters of critical point and gas-liquid coexistence for the models **IIs** and **Id** and parameters of triple point and gas-crystal coexistence in the model **Id** (Table I). Parameters of the phase transitions in all three models were found to be of the same order. Moreover some of these parameters in the models **Is** and **Id** almost coincide in reduced coordinates (Fig.1). At the same time the shape of coexistence curve for electron gas differs considerably from those for the mass-nonsymmetrical models **Is** and **Id** because of different behavior at $T \Rightarrow 0$ of specific heat of the classical OCP-crystal and the degenerated electron liquid. Note almost the same value (rather great in comparison with the real substances) of the ratio of normal density to the critical one for all the models.

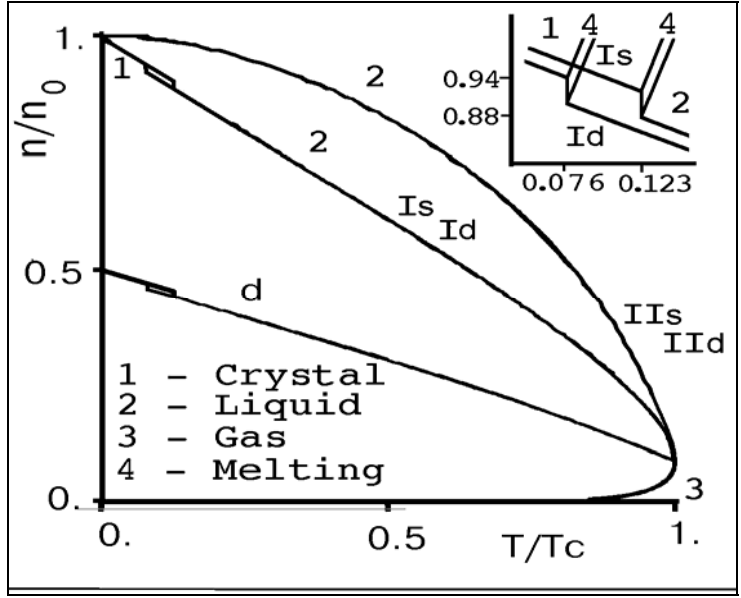

**Figure.1**. Two-phase coexistence in the Simple- and Double-OCP ($Z = 1$) and in the Electron gas (**Is**,**IIs**,**Id**) in reduced coordinates: $T_C$ - critical temperature, $n_0$ - normal density; $d$ - diameter of phase boundary in **Is**, **Id**. (On insertion - structure of the phase boundary in the vicinity of triple point).

**Table I**. Parameters of the phase transitions in the Electron gas and the Double-OCP for different values of charge number, $Z$: I - critical points; II - triple points.

|   | $Z$ (Double-OCP) | 1 | 2 | 3 | 10 | Electron gas |
|---|---|---|---|---|---|---|
| I | $T_C(Ry)$ | 0.100 | 0.283 | 0.508 | 2.69 | 0.0475 |
|   | $\Gamma_C$ | 5.54 | 9.43 | 13.2 | 39.4 | 4.54 |
|   | $(r_s)_C$ | 3.58 | 2.38 | 1.86 | 0.87 | 9.27 |
|   | $(n_e \Lambda_e^3)_C$ | 7.26 | 5.22 | 4.55 | 3.60 | 1.29 |
|   | $\{P/(n_i + n_e)kT\}_C$ | 0.107 | 0.111 | 0.115 | 0.122 | 0.0655 |
| II | $T_{tr}(Ry)$ | 0.00767 | 0.0333 | 0.0803 | 1.15 | - |
|   | $(\Delta n/n)_{tr}$ | 0.012 | 0.014 | 0.015 | 0.018 | - |

$$\Gamma \equiv (4\pi n/3)^{1/3}(Z^2 e^2/kT) \quad \Lambda_e^2 \equiv 2\pi\hbar^2/m_e kT \quad r_S^{-3} \equiv 4\pi n_e a_B^3/3 \quad a_B \equiv \hbar^2/m_e e^2$$

### 3. OCP phase transition and anomalies in the equilibrium non-uniform charge distribution.

The phenomenon which looks like phase transition in the OCP at macroscopic level manifests itself at appropriate conditions (low temperatures and densities) at microscopic level as a discontinuity in equilibrium non-uniform charge profile. It occurs when simple local approximations of Thomas-Fermi-Dirac or Poisson-Boltzmann-Debye-type are used. In present work we have calculated the charge distributions in the following cases:

- Electron profile in atomic cell [12]
- Electron profile near charged hard wall
- Electron profile near the border of semi-infinite rigid positive background [13].
- Counter-ion distribution around a polyion in electrolyte [14] or free ions distribution around charged condensed particle in dusty plasma.

We have used the density-functional theory description [12]

$$F[n(\cdot)] = Ze\int \varphi_{ext}(\bar{x})\cdot n(\bar{x})d\bar{x} + \frac{Z^2e^2}{2}\int \frac{n(\bar{x})\cdot n(\bar{y})}{|\bar{x}-\bar{y}|}d\bar{x}d\bar{y} + F^*[n(\cdot)] \qquad (1)$$

The free energy functional (1) was minimized over electron (or ion) density $n(r)$ with the additional normalization condition of electroneutrality. In the local ("*only on density*") approximation the exchange-correlation-kinetic term $F^*$ is:

$$F^*[n(\cdot)] = \int f(n(\bar{x}))\cdot n(\bar{x})d\bar{x}, \qquad f(n) = \lim \left\{\frac{F(N,V,T)}{N}\right\}_{N/V\to n}^{N\to\infty, V\to\infty} \qquad (2)$$

If we take $F(N,V,T)$ in (2) as the ideal gas free energy, we deal with the Thomas-Fermi or Poisson-Boltzmann approximations [12]. In present calculations we used in Eq.2 the exact free energy $F^{OCP}(N,V,T)$ of considered OCP of electrons (**IIs**) or ions (**IIIs**) on *uniformly compressible* background.

Everywhere we have obtained similar results. At sufficiently low temperatures (and densities) the predicted discontinuity appears in the charge profiles.

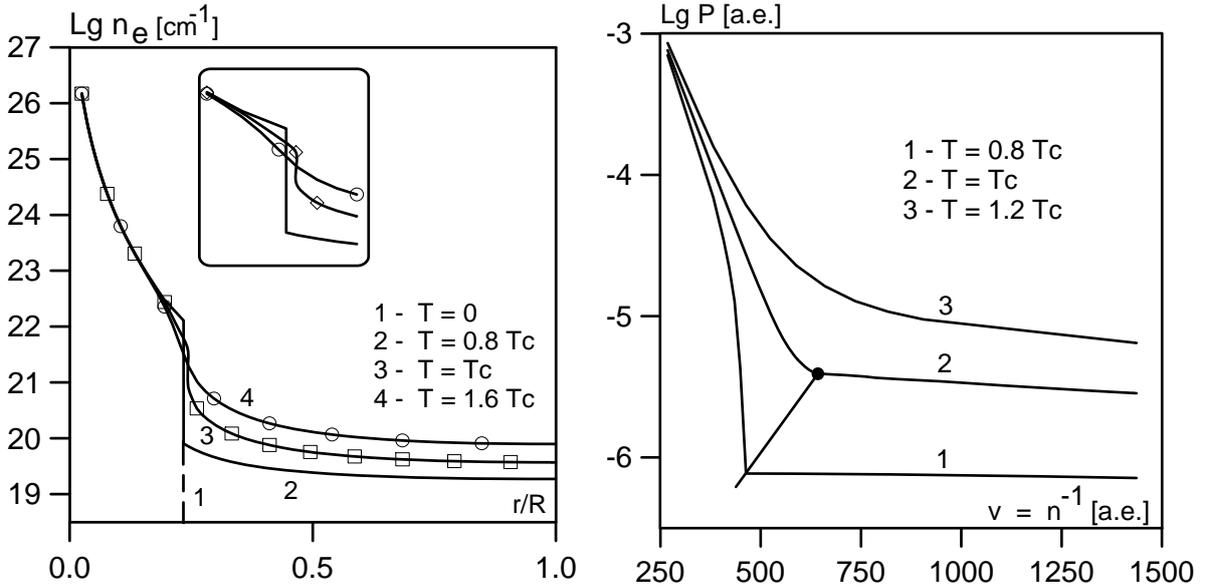

**Figure 2**. Electron distribution in Pb atomic cell calculated in frames of DFT scheme (1),(2) with the use of equation of state *f(n)* of interacting electron gas (**IIs**).

**Figure 3**. Electron contribution in equation of state of Pb, which was obtained with the use of calculated atomic cell electron distribution.

In terms of the OCP phase transition this boundary temperature is just its critical temperature $T_C$, and two values of the local densities at the discontinuity are the coexisting densities of "condensed" and "gaseous" phases, which depend only on temperature. For example, the electron profile in atomic cell, calculated with this approximation for

$T < T_C^{el.gas} \cong 0.65\ eV$ and $n \leq n_C^{el.gas}$, breaks up into condensed "drop" around the attractive center and diffuse "atmosphere" at the cell periphery (Fig.2). Thus we can predict similar decomposition at low temperature ($T < 0.65\ eV$) for the system of many nuclei embedded in electron fluid. For example, it must be so in the Thomas-Fermi Molecular Dynamic (TFMD) approach of Clerouin and Zerah [15] where the Thomas-Fermi-Dirac-type approximation has been used. Moreover, we may expect the decomposition of whole system of electrons and nuclei into two electroneutral phases of different densities.

Results of the calculations of electron distribution in atomic cell can be used for obtaining of total equation of state (EQS). It is well-known [12] that in the local approximation the electron contribution to the EQS depends only on boundary value of electron density. Thus, when the cell being compressed and its boundary touches the discontinuity, it results in corresponding anomaly in EQS. For $T < 0.65\ eV$ the break appears in isotherms and the discontinuity - in isothermal compressibility. Note that the locus of discussed singularities in $P \leftrightarrow T$ coordinates is completely equivalent to the saturation curve of the phase transition in considered model of electron gas with compressible background.

## Acknowledgements

Work supported by the Grant N 94-02-04-a of Russian Foundation of Basic Research.